\documentclass[aps,preprint,prd,nofootinbib]{revtex4-1}

\usepackage{graphicx}

\begin{document}
\title{A Comparative Study of $f_B$ within QCD Sum Rules with Two Typical Correlators up to Next-to-Leading Order}
\date{\today}
\author{Xing-Gang Wu \footnote{wuxg@cqu.edu.cn}, Yao Yu, Gu Chen and  Hua-Yong Han}
\address{Department of Physics, Chongqing University, Chongqing 400044, P.R. China}

\begin{abstract}

The $B$-meson decay constant $f_B$ is an important component for studying the $B$-meson decays, which can be studied through QCD sum rules. We make a detailed discussion on $f_B$ from two sum rules up to next-to-leading order, i.e. sum rules I and II, which are derived from the conventional correlator and the correlator with chiral currents respectively. It is found that these two sum rules are consistent with each other. The sum rules II involves less non-perturbative condensates as that of sum rules I, and in principle, it can be more accurate if we know the dimension-four gluon condensate well. It is found that $f_B$ decreases with the increment of $m_b$, and to compare with the Belle experimental data on $f_B$, both sum rules I and II prefer smaller pole $b$-quark mass, $m_b=4.68\pm0.07$ GeV. By varying all the input parameters within their reasonable regions and by adding all the uncertainties in quadrature, we obtain $f_B=172^{+23}_{-25}$ MeV for sum rules I and $f_B=214_{-34}^{+26}$ MeV for sum rules II. \\



\end{abstract}

\maketitle


The study of $B$ physics plays a fundamental role both in accurately testing the Standard Model and in search of new physics. For such purpose, it is very import to determine the $B$-meson transition matrix elements with high accuracy. Especially, as has already been emphasized in many works in the literature, the simplest matrix element $f_B$, which is defined as $\langle 0|\bar q\gamma_\mu\gamma_5 b|B(p)\rangle = i f_B p_\mu$ or $(m_b+m_q)\langle 0|\bar q i\gamma_5 b|B(p)\rangle =m_B^2 f_B$ is fundamental for the physics of the heavy-light quark systems, where $q$ denotes the light $u$ or $d$ quark under the isospin symmetry. The decay constants usually are studied through their leptonic decays, earlier discussions of which can be found in Ref.\cite{khlopov}. As for $f_B$, its first direct measurement is done by Belle experiment, which shows that $f_B=229^{+36}_{-31}({\rm stat}) ^{+34}_{-37}({\rm syst})$ MeV from the measurement of the decay $B^{-}\to\tau\bar\nu_\tau$ \cite{exp}. At the present, the lattice QCD calculation and the QCD sum rules are two suitable approaches for theoretically extracting these elements and their results are usually complementary to each other. Some typical extractions of $f_B$ were investigated in Refs.\cite{sr1,sr2,sr3,sr4,sr5,sr6,sr7,sr8,sr9} from QCD sum rules and in Refs.\cite{lat1,lat2,lat3,lat4,lat5,lat6} from lattice calculation.

The QCD sum rules approach was developed by Shifman, Vainshtein and Zakharov (SVZ) in the seventies of last century \cite{svz}, and now, it becomes a widely adopted tool for studying hadron phenomenology. It was designed to determine the properties of the ground-state hadrons at low momentum transfer to the region of large momentum transfer. More explicitly, hadrons are represented by certain interpolating quark currents taken at large virtuality, and then the correlation function (correlator) of these currents is introduced and treated in the framework of operator product expansion (OPE) such that the short- and the long- distance quark-gluon interactions are separated. The short-distance interaction is calculated using perturbative QCD, while the long-distance one can be parameterized in terms of the universal non-perturbative vacuum condensates. Next, the result of QCD calculation is matched, via dispersion relation, to a sum over hadronic states. Finally, the Borel transformation is introduced to suppress or even cut off the irrelevant or unknown excited and continuum states so as to improve the accuracy of the obtained sum rules, since little is known about the spectral function of the excited and continuum states. In such a way, the so called SVZ sum rules combines the concepts of OPE, the dispersive representation of correlator and the quark-hadron duality consistently that allows the derived non-excited hadron states' properties to be within their reasonable theoretical uncertainties.

It is noted that one of the basic quantity for the QCD sum rules is the correlator. How to ``design" a proper correlator for a particular case is a tricky problem. By a suitable choice of the correlator, one can not only obtain the right properties of the hadrons but also simplify the theoretical uncertainties effectively. Usually the currents adopted in the correlators are taken to be those with definite quantum numbers, such as those with definite $J^P$, where $J$ is the total angular momentum and $P$ is the parity. However such kind of correlator is not the only choice adopted in the literature, e.g. the correlator constructed with chiral current is also adopted, which is firstly proposed by Ref.\cite{huang0} to study the process $B\to K^{*} \gamma$ and then developed in dealing with various processes \cite{chiral1,chiral2,chiral3,chiral4,huang1,huang2,wu}. For different correlators, we need to check whether the theoretical estimations are consistent with each other or not, or to determine which correlator can lead to more accurate or more appropriate estimation for a particular physical process. And it is one of the purpose of the present paper, to make a comparative study of $f_B$ under two typical choices for the two-point correlator.

The first correlator is the conventional one and is defined as
\begin{equation}
\Pi(q^2) = i\int d^4x e^{iqx}\langle0|
T\big\{\overline{q}(x)\gamma_5 b(x), \overline{b}(0)\gamma_5
q(0)\big\}|0\rangle \;, \label{cor1}
\end{equation}
and the second one is the correlator with chiral current,
\begin{equation}
\Pi(q^2)=i\int d^4x e^{iqx}\langle0|T\big\{\overline{q}(x)(1+\gamma
_5)b(x),\overline{b}(0)(1-\gamma_5)q(0)\big\}|0\rangle \;.
\label{cor2}
\end{equation}

Following the standard procedure of SVZ sum rules, we can obtain two sum rules for $f_B$ from the above defined correlators.

The sum rules up to next-to-leading order (NLO) for the first correlator can be written as (sum rules I)
\begin{widetext}
\begin{eqnarray}
f_B^2\frac{m_B^4}{m_b^2}e^{-m_B^2/M^2} &=&
\frac{3}{8\pi^2}\int\limits_{m_b^2} ^{s_0}ds s e^{-s/M^2}(1-x)^2
\Bigg [1 + \frac{\alpha_s(\mu_{IR}) C_F}{\pi}\rho(x)\Bigg ] \nonumber \\
&& +e^{-m_b^2/M^2}\Bigg[-m_b \langle \bar{q} q \rangle +
\frac{1}{12}\langle \frac{\alpha_s}{\pi} G G \rangle
-\frac{1}{2M^2}\left(1-\frac{m_b^2}{2M^2}\right)m_b \langle
\bar{q}g\sigma\cdot G q\rangle\nonumber\\
&& -\frac{16\pi}{27} \frac{\alpha_s(\mu_{IR})\langle \bar{q} q
\rangle^2}{M^2} \left(1-\frac{m_b^2}{4M^2} -\frac{m_b^4}
{12M^4}\right) \Bigg]~, \label{sr1}
\end{eqnarray}
\end{widetext}
where $m_b$ stands for the pole quark mass, $\mu_{IR}$ is the renormalization scale, $x=m_b^2/s$ and $C_F=4/3$. $M$ and $s_0$ stand for the Borel parameter and the effective continuum threshold respectively. Since $m_q$ is quite small in comparison to $m_b$ or $\mu_{IR}$, so terms proportional to $m_q$ have implicitly neglected. The function $\rho(x)$ determines the spectral density of the NLO correction to the perturbative part and it takes the following form
\begin{widetext}
\begin{equation}
\rho(x)=\frac{9}{4}+2{\rm Li}_2(x)+ \ln x \ln (1-x)-\ln(1-x)+\left(x-\frac{3}{2}\right) \ln\frac{1-x}{x}-\frac{x}{1-x}\ln x \;,
\end{equation}
\end{widetext}
where the dilogarithm function ${\rm Li}_2(x)=-\int_0^x\frac{dt}{t}\ln(1-t)$. Practically, $\rho(x)$ is firstly derived under the $\overline{MS}$ scheme, and then it is transformed to be the present form with the help of the well-known one loop formula for the relation between the $\overline{MS}$ $b$-quark mass and the pole quark mass, i.e. $\bar{m}_b(\mu_{IR})=m_b\left[1+ \frac{ \alpha_s(\mu_{IR}) C_F}{4\pi} \left(-4+3\ln\frac{m_b^2} {\mu_{IR}^2}\right)\right]$. In some references \cite{sr6,runmb}, it is argued that one should use $b$-quark running mass other than its pole quark mass to do the numerical analysis. However, we think these two choices are equivalent to each other in principle, since we need to use their relation either to change the bound state part (to be determined by the pole quark mass) with the running mass or to change the hard scattering part (to be calculated by the running mass) with the pole quark mass. Moreover, as a cross check of the present obtained formulae, it is found that the sum rules I agrees with those of Refs.\cite{sr1,sr2,sr6,sr7,sr8} when taking the same approximation and the same type of $b$-quark mass.

The sum rules up to NLO for the second correlator can be written as (sum rules II):
\begin{widetext}
\begin{eqnarray}
f_B^2 \frac{m_B^4}{m_b^2}e^{-m_B^2/M^2}&=&
\frac{3}{4\pi^2}\int\limits_{m_b^2} ^{s'_0}ds s e^{-s/M^2}(1-x)^2
\Bigg [1 + \frac{\alpha_s(\mu_{IR}) C_F}{\pi}\rho(x)\Bigg ]
\nonumber\\
&& +e^{-m_b^2/M^2}\Bigg[\frac{1}{6}\langle \frac{\alpha_s}{\pi} G G
\rangle -\frac{32\pi}{27} \frac{\alpha_s(\mu_{IR})\langle \bar q q
\rangle^2}{M^2} \left(1-\frac{m_b^2}{4M^2}
-\frac{m_b^4}{12M^4}\right) \Bigg] \;,\label{sr2}
\end{eqnarray}
\end{widetext}
where $s'_0$ also is the effective threshold. Again, the terms proportional to $m_q$ are neglected due to their smallness. It can be found that the sum rules II is simpler in form than that of sum rules I.

As a comparison of sum rules I and II, it can be found that by taking proper chiral current in the correlator, one can reduce the theoretical uncertainties to a certain degree. Especially, in sum rules I, the first non-perturbative term and hence the dominant contribution of the non-perturbative condensates is the dimension-three quark condensate $\langle\bar{q}q\rangle$. And, in sum rules II, the first non-perturbative term is the dimension-four gluon condensate $\langle \frac{\alpha_s}{\pi} G G \rangle$. Therefore, with proper chiral currents in the correlator, one can naturally suppress the non-perturbative sources and hence improve the accuracy of the obtained sum rules. This observation is similar to the case of light-cone sum rules for $B\to $ pseudo-scalar transition form factors where with chiral current in the correlator, one can eliminate the most uncertain twist-3 contributions \cite{huang1, huang2, wu} in comparison to the sum rules derived by using the conventional correlator (see Ref.\cite{pball} for an explicit example).

Schematically, the two sum rules I and II can be rewritten in the following simpler form
\begin{equation} \label{sumfb}
f_B^2 \frac{m_B^4}{m^2_b} e^{-m^2_B/M^2} =\int^{s_0}_{m_b^2}
\rho^{tot}(s) e^{-s/M^2}ds ,
\end{equation}
where the spectral density $\rho^{tot}(s)$ can be read from Eqs.(\ref{sr1},\ref{sr2}). It is noted that the continuum threshold $s_0$ should be varied with the different choice of correlators. While more practically, the Borel parameter $M^2$ and the continuum threshold $s_0$ are determined in such a combined way that the resulting $B$-meson decay constant $f_B$ does not depend too much on the precise values of these parameters. In addition, the following three criteria are adopted for numerically determining the range of of $(M^2,s_0)$:
\begin{itemize}
\item Criterion (A): The continuum contribution, that is the part of the dispersive integral from $s_0$ to $\infty$, should not be too large, e.g. less than $30\%$ of the total dispersive integral. As will be found later that this condition is well satisfied, since the resultant ranges for $(M^2,s_0)$ shall led this ratio be less than 20\% for sum rules I and be less than 15\% for sum rules II.
\item Criterion (B): The contributions from the dimension-six condensate terms shall not exceed $15\%$ for $f_B$.
\item Criterion (C): The derivative of the logarithm of Eq.(\ref{sumfb}) with respect to $(-1/M^2)$ gives the $B$-meson mass $m_B$, i.e. $ m_B^2=\int^{s_0}_{m_b^2}\rho^{tot}(s)e^{-s/M^2} s ds {\Bigg /} \int^{s_0}_{m_b^2}\rho^{tot}(s) e^{-s/M^2}ds$, and we require its value to be full-filled with high accuracy, i.e. $\lesssim 0.1\%$.
\end{itemize}
Here, we shall not give any further constraints on $s_0$, and we only treat it as an effective scale, which is roughly around the squared mass of $B$-meson first excited state.

In numerical evaluation of the sum rules I and II, we take $m_B=5.279$ GeV and use the  following values for the condensates \cite{duplan,cond}:
\begin{eqnarray}
\langle\bar{q}q\rangle(1{\rm GeV}) &=& -(0.246^{+0.018}_{-0.019}{\rm GeV})^3, \nonumber\\
\langle \frac{\alpha_s}{\pi} G G \rangle &=& 0.012^{+0.006}_{-0.012}{\rm GeV}^4 ,\nonumber\\
\langle \bar{q}g\sigma\cdot G q\rangle(1{\rm GeV}) &=& (0.8\pm0.2)
GeV^2\langle\bar{q}q\rangle(1{\rm GeV}). \label{parameter}
\end{eqnarray}
It is found that the relative error for $\langle \frac{\alpha_s}{\pi} G G \rangle$ is quite large. Further more, we shall use LO anomalous dimensions of the condensates to evaluate them up to the renormalization scale, $\mu_{IR}=\sqrt{m_B^2-m_b^2}$ \cite{pball}. The $b$-quark pole mass is taken as: $m_b\simeq 4.68 ^{+0.17}_{-0.07}$ GeV, which is the recent world average suggested by Particle Data Group \cite{mbmass}. As for the running coupling constant, we approximate it by a one-loop form, $\alpha_s(\mu_{IR})=4\pi/(9\ln(\mu^2_{IR} /\Lambda^2_{QCD}))$, where $\Lambda_{QCD}=0.241$ GeV that is determined by reproducing $\alpha_s(M_z)=0.1176$ \cite{mbmass}.

Criteria (A, B, C) determine a set of parameters for each value of $m_b$. In the following, we shall firstly take $m_b=4.68$ GeV as an explicit example to show how to determine the allowable ranges for $M^2$ and $s_0$ and to derive $f_B$ within these ranges. And then we shall present the results for several typical $m_b$.

\begin{figure}
\includegraphics[scale=0.5]{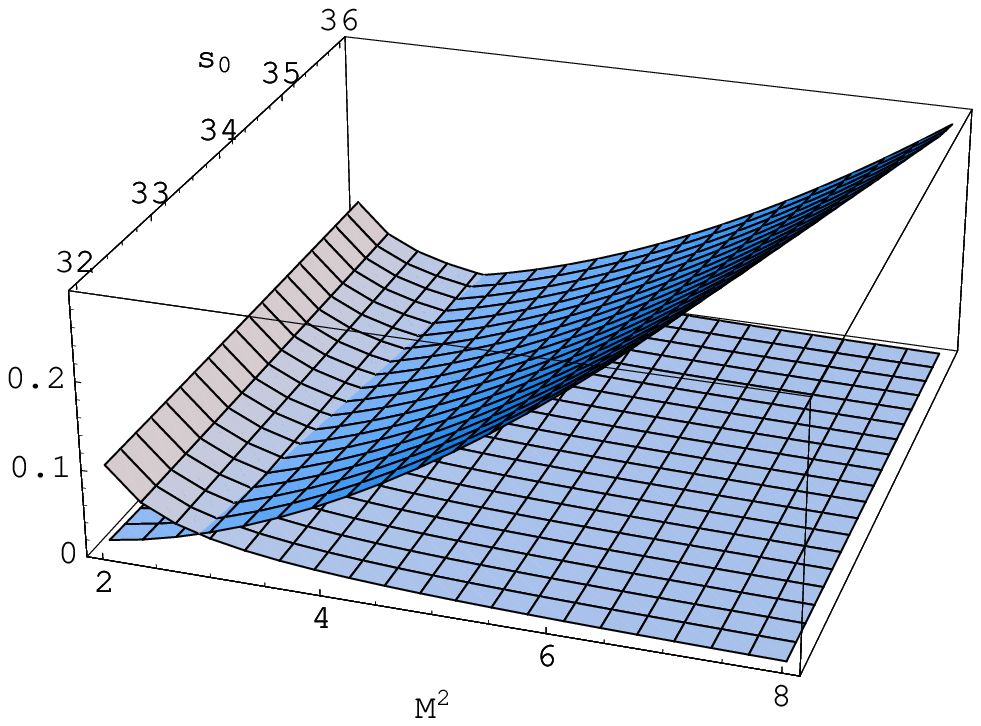}
\includegraphics[scale=0.55]{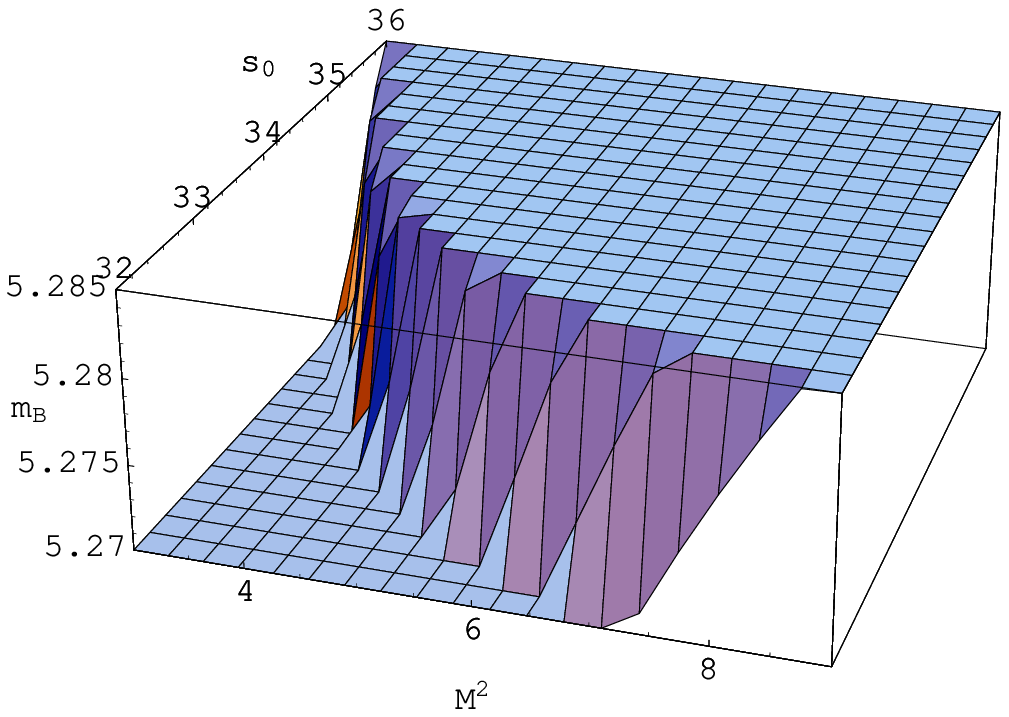}
\caption{Allowable range of $(M^2,s_0)$ for sum rules I, where $m_b=4.68$ GeV and the non-perturbative condensates are set to be their center values. The left diagram is for criteria (A) and (B), whose third axis is for the ratio of the continuum contribution and the dimension-six contribution over the total contributions, respectively. The right diagram is for criterion (C).} \label{fig1}
\end{figure}

\begin{figure}
\includegraphics[scale=0.5]{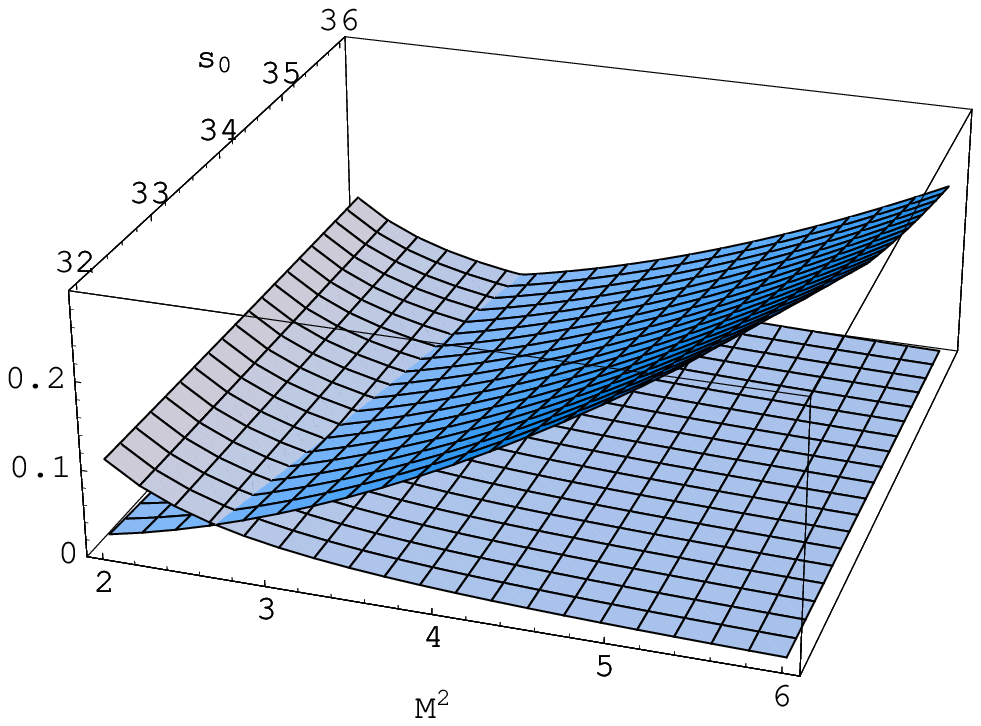}
\includegraphics[scale=0.55]{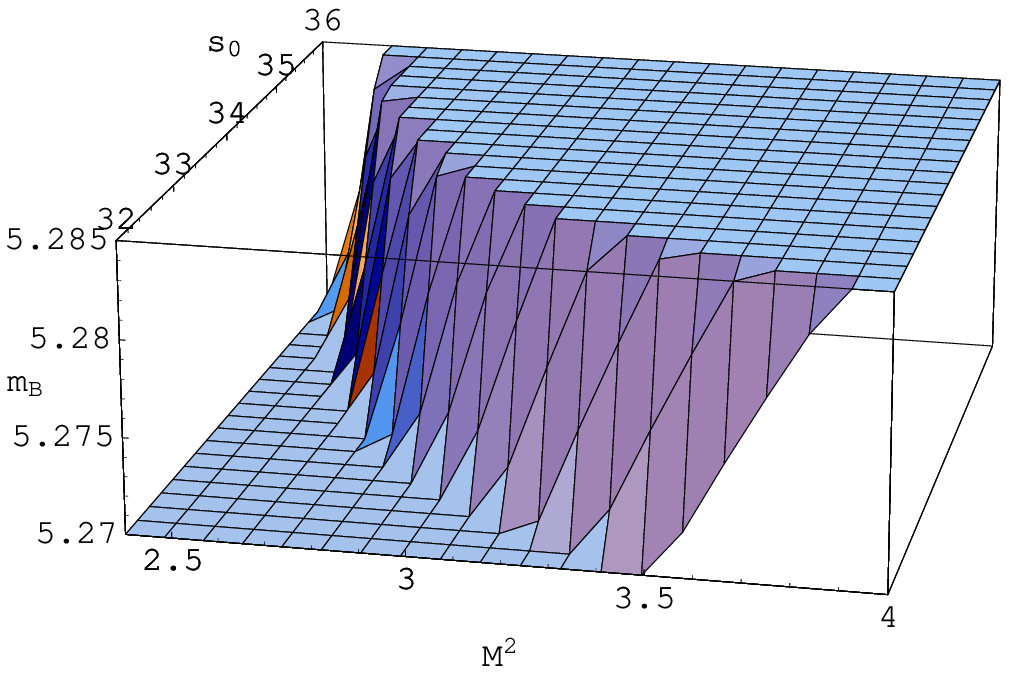}
\caption{Allowable range of $(M^2,s_0)$ for sum rules II, where $m_b=4.68$ GeV and the non-perturbative condensates are set to be their center values. The left diagram is for criteria (A) and (B), whose third axis is for the ratio of the continuum contribution and the dimension-six contribution over the total contributions, respectively. The right diagram is for criterion (C). } \label{fig3}
\end{figure}

\begin{figure}
\includegraphics[scale=0.3]{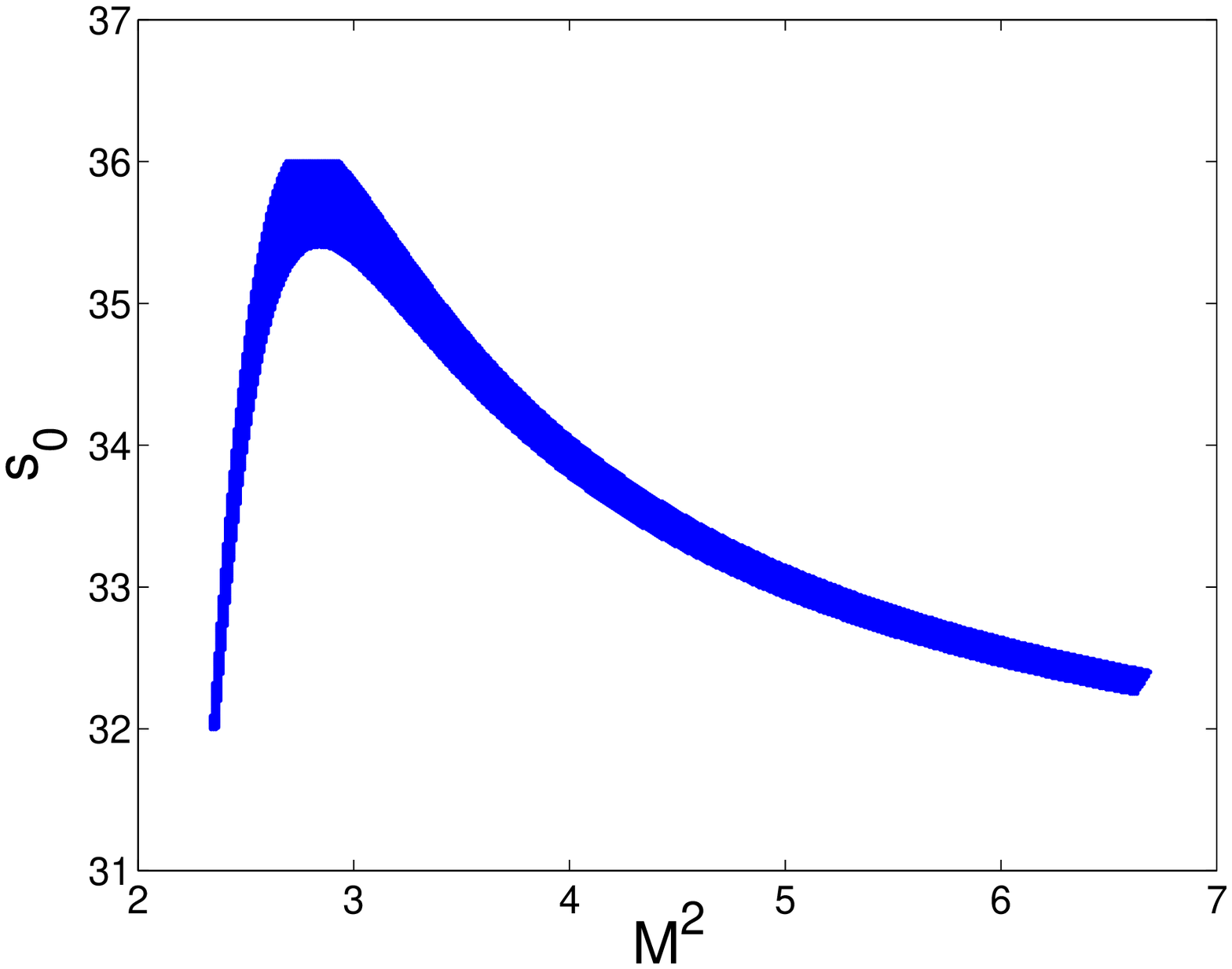}
\includegraphics[scale=0.3]{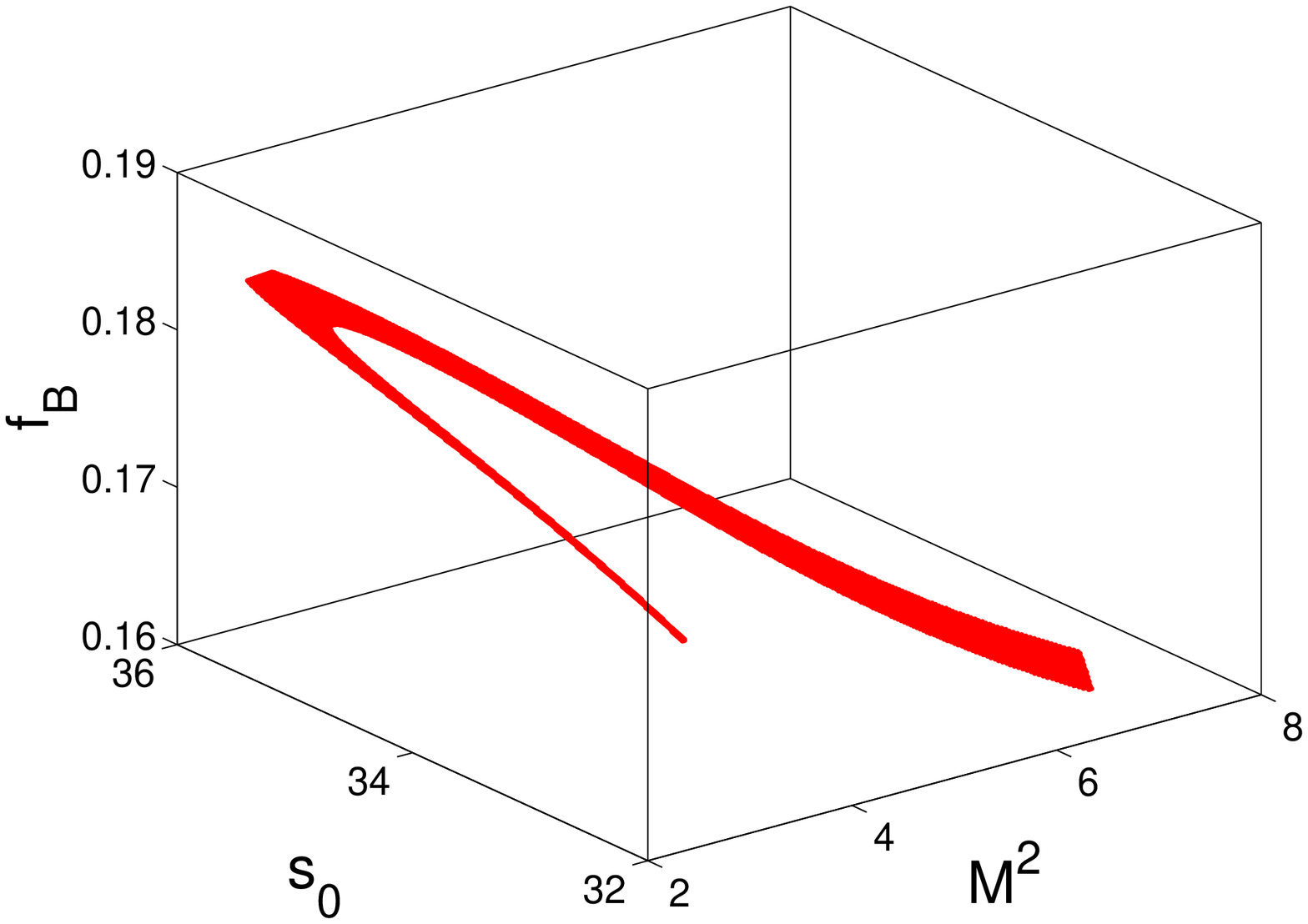}
\caption{The allowable range of $(M^2,s_0)$ for sum rules I (Left diagram) and the value of $f_B$ within this range (Right diagram), where $m_b=4.68$ GeV and the non-perturbative condensates are set to be their center values.  }\label{fig2}
\end{figure}

\begin{figure}
\includegraphics[scale=0.3]{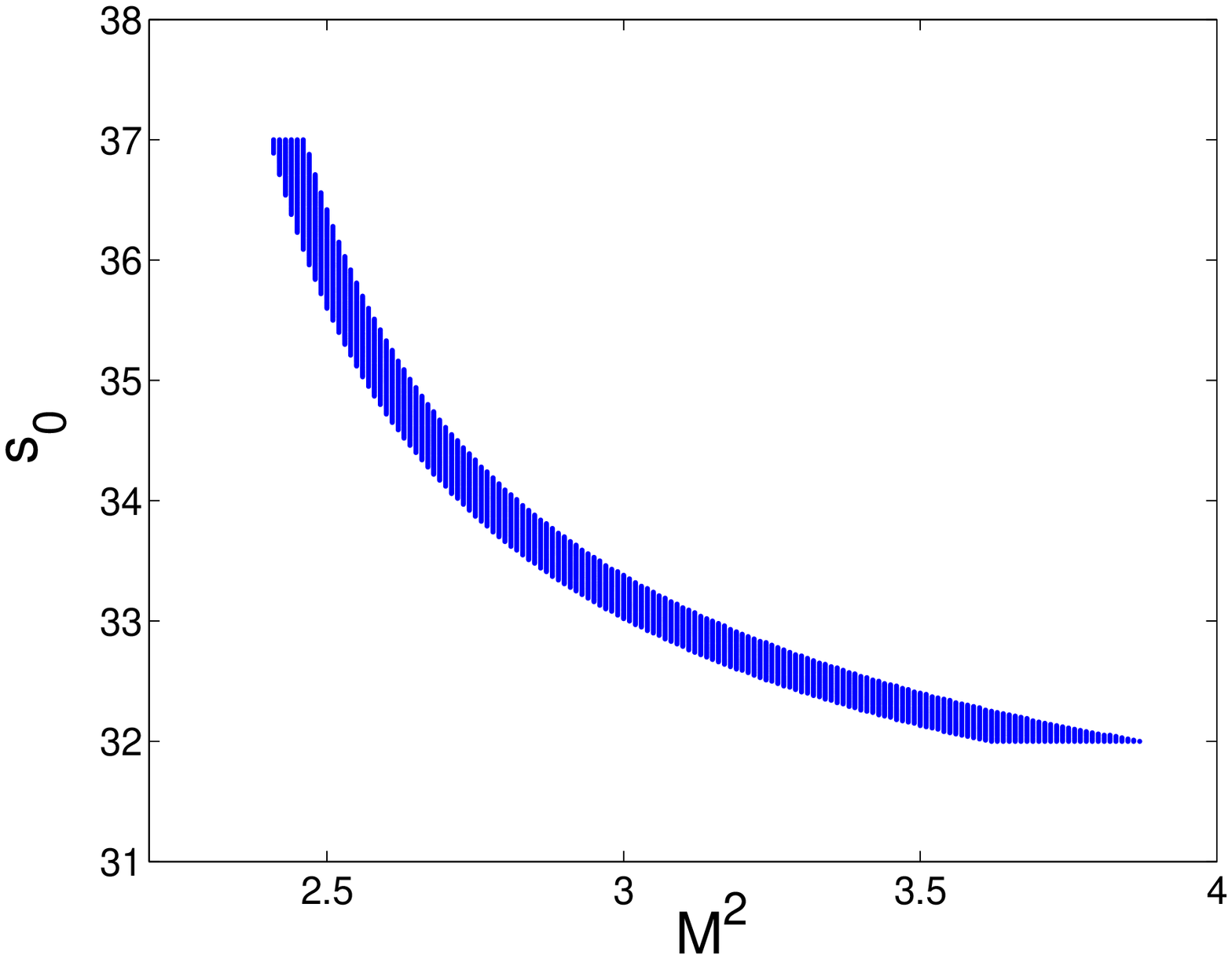}
\includegraphics[scale=0.3]{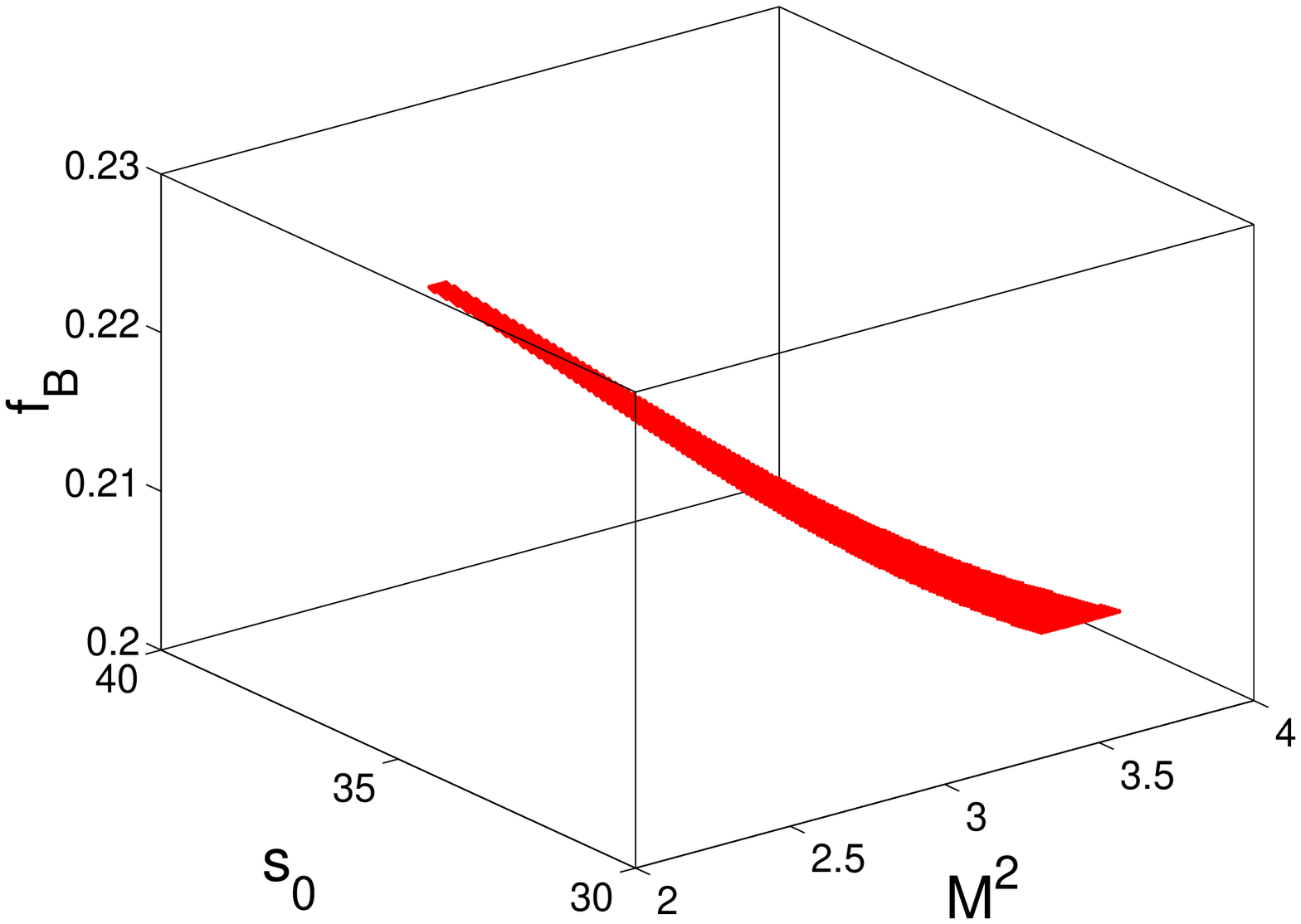}
\caption{The allowable range of $(M^2,s_0)$ for sum rules II (Left diagram) and the value of $f_B$ within this range (Right diagram), where $m_b=4.68$ GeV and the non-perturbative condensates are set to be their center values. } \label{fig4}
\end{figure}

Firstly, the three-dimensional figures (\ref{fig1},\ref{fig3}) present the allowable range of $M^2$ and $s_0$ for sum rules I and II. The left diagram of Figs.(\ref{fig1},\ref{fig3}) is for criteria (A) and (B), whose first and second axes are for $M^2$ and $s_0$, and third axis is for the ratio of the continuum contribution over the total contribution for criteria (A) and the ratio of the dimension-six contribution over the total contribution for criteria (B) respectively. By setting a fixed value for $s_0$, it is found that the continue contribution increases with the {\it increment} of $M^2$ and the contribution from the dimension-six condensate term increases with the {\it decrement} of $M^2$, so criterion (A) determines the upper limit of $M^2$ and criterion (B) determines the lower limit of $M^2$. Then, a possible Borel window can be obtained from criteria (A) and (B), which shall be further constrained by criterion (C). Practically, the range of $(M^2,s_0)$ can be determined numerically. The left diagram of Fig.(\ref{fig2}) shows the allowable range of $(M^2,s_0)$ for sum rules I, which is obtained by sampling $500\times400$ points within $M^2\in [2,7] \;{\rm GeV}^2$ and $s_0\in[32,36]\; {\rm GeV}^2$. And the left diagram of Fig.(\ref{fig4}) shows the allowable range of $(M^2,s_0)$ for sum rules II, which is obtained by sampling $400\times500$ points within $M^2\in [2,6]\; {\rm GeV}^2$ and $s_0\in[32,37] \;{\rm GeV}^2$. The right diagrams of Figs.(\ref{fig2},\ref{fig4}) present the corresponding values of $f_B$ within its allowable region of $M^2$ and $s_0$. For fixed $M^2$, it can be found that $f_B$ increases with the increment of $s_0$. And for fixed $s_0$, $f_B$ shall be steady under the reasonable region of the Borel parameter $M^2$, i.e. the uncertainty is less than $3\%$. By varying $(M^2,s_0)$ within their allowable region, the uncertainties of $f_B$ is less than $6\%$ for sum rules I and less than $5\%$ for sum rules II. Such small uncertainty for a fixed $m_b$ agrees with the requirement of the reliability of sum rules.

\begin{table}
\centering
\begin{tabular}{|c||c|c|c|}
\hline ~~~ - ~~~ & ~$s_0 (GeV^2)$~ & ~$M^2 (GeV^2)$~ & ~~~ $f_B (MeV)$ ~~~\\
\hline\hline
~~$m_b=4.85$~~ & 31.7 & 4.88 & $119\pm6$  \\
\hline
~~$m_b=4.75$~~ &  33.3 & 3.52  &  $149\pm8$ \\
\hline
$m_b=4.68$ & 35.9  & 2.99  & $172\pm10$  \\
\hline
$m_b=4.61$ & 36.9  & 2.25  & $193\pm15$  \\
\hline
\end{tabular}
\caption{Sum rules I for $f_B$, where the values of $s_0$ and $M^2$ are those that lead to the maximum value of $f_B$ for a particular $m_b$. The errors are caused by  varying $(M^2,s_0)$
within the region that is determined by setting the non-perturbative condensates to be their center values. } \label{tabfbI}
\end{table}

\begin{table}
\centering
\begin{tabular}{|c||c|c|c|}
\hline ~~~ - ~~~ & ~$s_0 (GeV^2)$~ & ~$M^2 (GeV^2)$~ & ~~~ $f_B (MeV)$ ~~~\\
\hline\hline
~~$m_b=4.85$~~ & 33.2  & 2.00  & $139\pm12$  \\
\hline
~~$m_b=4.75$~~ & 35.9  & 2.16  &  $181\pm14$ \\
\hline
$m_b=4.68$  &  36.9 & 2.41  & $214\pm10$  \\
\hline
$m_b=4.61$  & 38.9 & 2.63 & $238\pm17$  \\
\hline
\end{tabular}
\caption{Sum rules II for $f_B$, where the values of $s_0$ and $M^2$ are those that lead to the maximum value of $f_B$ for a particular $m_b$. The errors are caused by varying $(M^2,s_0)$ within the region that is determined by setting the non-perturbative condensates to be their center values. } \label{tabfbII}
\end{table}

Secondly, we present $f_B$ for several typical $m_b$, i.e. $m_b$=4.61 GeV, 4.68 GeV, 4.75 GeV and 4.85 GeV, respectively. The allowable region of $(M^2,s_0)$ for each $m_b$ can be determined through the same procedure described above. The results are put in TABs.(\ref{tabfbI},\ref{tabfbII}), where the listed values of $s_0$ and $M^2$ are those that lead to the maximum value of $f_B$ for a particular $m_b$. And the listed errors for $f_B$ are determined by setting the non-perturbative condensates to be their center values. By further varying the non-perturbative condensates within the region of Eq.(\ref{parameter}), an extra $\pm1$ MeV up to $\pm4$ MeV error should be taken into consideration. For examples, by setting $m_b=4.68$ GeV, an extra $\pm2$ MeV error shall be caused by those non-perturbative condensates for both sum rules I and II; while by setting $m_b=4.61$ GeV, it changes to be $\pm1$ MeV. Furthermore, it can be found that the value of $f_B$ decreases with the increment of $m_b$, and to compare with the experimental data on $f_B$, both sum rules I and II prefer small $b$-quark pole mass, i.e. $m_b=4.68\pm0.07$ GeV. It is found that the main uncertainty comes from $m_b$, so a better understanding of $m_b$ shall further improve the present sum rules. For example, in Ref.\cite{duplan}, a smaller region of $m_b\in[4.55, 4.60]$ GeV is adopted, which can be obtained by translating the running $b$-quark mass used there to the present pole mass, and then their result for $f_B$ is $214^{+7}_{-5}$ MeV. It should be noted that such a small error is also due to a smaller Borel window and a smaller region of $s_0$ derived there \cite{duplan}. However if we treat $(M^2,s_0)$ as correlated parameters then their separate region shall be broadened as shown by the left diagrams of Figs.(\ref{fig2},\ref{fig4}).

As a summary, we have presented a detailed discussion on $f_B$ from two sum rules, i.e. sum rules I and II, which are derived from the conventional correlator and the correlator with chiral currents respectively. Under proper parameter values, both sum rules I and II can lead to reasonable $f_B$ that is consistent with the Belle experiment. At a fixed $m_b$, the estimated uncertainties of $f_B$ from both sum rules I and II are less than $10\%$. If further fixing $s_0$, $f_B$ shall be quite steady versus the Borel parameter $M^2$. Hence the contributions from the uncertainty sources, e.g. the continuum states and higher dimensional condensates, are well under control. With the help of TABs.(\ref{tabfbI},\ref{tabfbII}), by varying $m_b$ within a small region $m_b=4.68\pm0.07$ GeV, we obtain $f_B=172^{+23}_{-25}$ MeV for sum rules I and $f_B=214_{-34}^{+26}$ MeV for sum rules II, where the errors from all the mentioned parameters such as $m_b$, $(M^2,s_0)$ and the non-perturbative condensates are added together in quadrature. Furthermore, the comparison of the sum rules I and II shows that we can improve the QCD sum rules in principle by using chiral current in the correlator. At the present, due to the large uncertainty of the dimension-four gluon condensates, such improvement on the sum rules with chiral current is not so clear as that of QCD light-cone sum rules for $B\to P$ transition form factors \cite{huang1, huang2, wu}. However with a more accurate values for these condensates, sum rules II shall become more accurate.

\hspace{2cm}

Acknowledgements: This work was supported in part by Natural Science Foundation Project of CQ CSTC under Grant No.2008BB0298, by Natural Science Foundation of China under Grant No.10805082 and No.11075225, and by the Fundamental Research Funds for the Central Universities under Grant No.CDJZR101000616.

\end{document}